\documentclass[twocolumn,showpacs,preprintnumbers,amsmath,amssymb]{revtex4}
\pdfoutput=1

\usepackage{graphicx}
\usepackage{dcolumn}
\usepackage{bm}

\usepackage[latin1]{inputenc}
\usepackage{subfigure}

\begin{document}


\title{Slowing light through Zeeman Coherence Oscillations in a duplicated two-level system }

\author{F.A. Hashmi}
\author{M.A. Bouchene}%
 \email{aziz@irsamc.ups-tlse.fr}
\affiliation{%
Laboratoire de Collisions Agrégats Réactivité, C.N.R.S. UMR 5589, IRSAMC\\
Université Paul Sabatier, 118 Route de Narbonne, 31062 TOULOUSE CEDEX 4, FRANCE
}%

\date{\today}

\begin{abstract}
We present the theory of a new method to slow a linearly polarized probe pulse as it propagates through a duplicated two-level system driven by an orthogonally polarized control field. The method makes use of Zeeman coherence oscillations (CZO) that arise in the atomic system because of the spatial and temporal modulation of the total polarization. This method exhibits properties similar to those that rely on the electromagnetically induced transparency (EIT) but \textit{without the existence of any trapping dark state}. We demonstrate also the propagation of a polariton in the medium.
\end{abstract}

\maketitle
\newcommand{\fpi}{\Omega_{\pi}}
\newcommand{\fsig}{\Omega_{\sigma}}
\newcommand{\rpi}{\rho_{\pi}}
\newcommand{\rpis}{\rho_{\pi}^{*}}
\newcommand{\rsig}{\rho_{\sigma}}
\newcommand{\rsigs}{\rho_{\sigma}^{*}}
\newcommand{\rsigm}{\rho_{\sigma}^{\left(1\right)}}
\newcommand{\rzg}{\rho_{zg}}
\newcommand{\rze}{\rho_{ze}}
\newcommand{\phsm}{e^{-i\Phi\left(\vec{r},t\right)}}
\newcommand{\delp}{\Delta_{\pi}}
\newcommand{\delpb}{\bar{\Delta}_{\pi}}

\newcommand{\delb}{\underline{\Delta}}

\newcommand{\gmm}{\Gamma}
\newcommand{\gmmd}{\Gamma_d}
\newcommand{\gmmz}{\Gamma_{ze}}
\newcommand{\nge}{n_g^{\left(0\right)}-n_e^{\left(0\right)}}
\newcommand{\rzo}{\rho_{zg}^{\left(1\right)}+\rho_{ze}^{\left(1\right)}}

The possibility to slow down light through quantum interferences is one of the most exciting research fields in physics. The key idea is to create a narrow transparency window in the absorption spectrum of a pulse as it propagates in a medium. This transparency window is related, through Kramers-Krönig relations to an abrupt variation of the refractive index $n(\omega)$ of the material leading to a small propagation velocity of the pulse $v_g=c\left( n(\omega)+\omega dn(\omega)/d\omega\right)^{-1}$.  The first realization of slow light \cite{ref1} made use of EIT \cite{ref2}. This effect can be realized in a three-level $\Lambda$ system excited by two fields on different transitions. The two lower states of the system give rise to a bright and a dark state where the dark state is immune to field excitation. Under suitable conditions of excitation, all the atoms can be coherently trapped in the dark state rendering the system transparent to both fields. This is Coherent Population Trapping (CPT) and it plays an important role in many phenomena in optics \cite{ref3}. Another technique to slow light makes use of Coherent Population Oscillations (CPO) \cite{ref4}. A two level system excited by a pump and a probe field having slightly different frequencies gives rise to population oscillations at the beat frequency inducing a population grating. Self diffraction of the pump from this grating into the probe field compensates for the absorption of the latter by the medium and hence gives rise to a transparency window. These two methods produce very narrow transparency windows leading to ultra-slow lights. These methods have been used to even stop, store light and to make optical quantum memories \cite{ref5}.  Many alternative techniques have also been proposed to obtain optically controllable delays for telecommunication purposes \cite{ref6} but light velocities achieved here are far from those obtained with EIT and CPO methods.

\begin{figure}
\includegraphics[width=3.5 in]{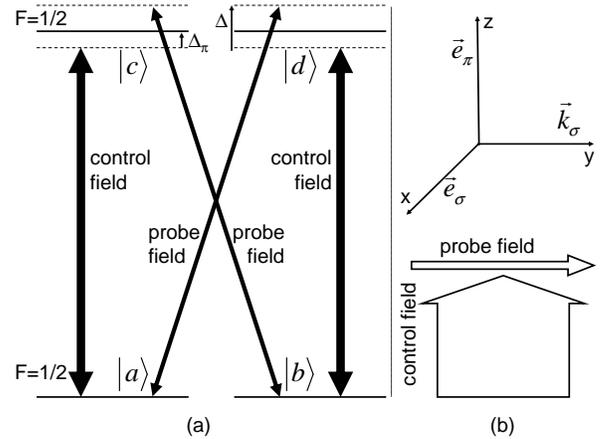}
\caption{\label{fig1} (a) The duplicated two-level system and (b) fields configurations.}
\end{figure}

In this letter, we present the theory of a new method that can lead to very small group velocities for a probe pulse that propagates in a duplicated two-level system driven by a control field (Fig.\ \ref{fig1}). This system can be considered as the degenerated version of the double $\Lambda$ systems that have been investigated in the context of resonant nonlinear optical phenomena \cite{ref7}. These systems can exhibit EIT phenomena for equal strength pulses having special phase relation between them \cite{ref8}. For these matching conditions, a dark state arises in the system, leading to CPT and hence to slow \cite{ref9} and stored \cite{ref10,ref11} light. However, for arbitrary field strengths and/or arbitrary phase relations between the fields, there is no dark state in the system, ruling out the possibility of transparency window or slow light through CPT. We show in this letter that we can still produce transparency in the system by making the fields propagate in different directions and by having the control field much stronger than the probe. The space/time modulation of the total polarization induces a grating in the zeeman coherences (\textit{Zeeman Coherence Oscillations} CZO) which diffracts the control field into the probe field compensating for the absorption of the latter. This is in strong analogy with CPO in a \textit{non collinear geometry} \cite{ref12} but in this case, it's the coherence that is oscillating and not the population. Moreover the transparency obtained exhibits properties more similar to the one obtained by EIT than CPO. In a previous study of the same system in the femtosecond regime, we have shown that coherent control of the medium gain can be achieved for the probe pulse \cite{ref13}. The present study is valid only for long pulses having duration larger than inverse relaxation rates.

Consider $F_{1/2}\rightarrow F_{1/2}$ transition (energy $\hbar\omega_0$) excited by two orthogonally polarized fields that propagate at right angle to each other (e.g. $^2S_{1/2}F_{1/2}\rightarrow ^2P_{1/2}F_{1/2}$ transition of $^6Li$ at 671 nm). The electric fields are $\vec{E_i}\left(\vec{r},t\right)=\vec{e_i}\epsilon_i\left(\vec{r,t}\right)e^{-i\left(\omega_i t-\vec{k_i}\cdot\vec{r}\right)}$ with $i=\sigma,\pi$. The system (Fig.\ \ref{fig1}) is equivalent to a duplicated two-level system with the $\pi$ polarized control field $\vec{E_{\pi}}$ coupling the transitions with identical $m_F$ and the $\sigma$ polarized probe field $\vec{E_{\sigma}}$ coupling the levels with different $m_F$. The control field is much stronger than the probe and we are interested in the spectral response of the atomic system for the probe pulse. Using the distributed pump configuration (Fig.\ \ref{fig1}) serves two purposes. First, in our method transparency is not achieved for the control field in contrast with EIT. Making the control beam cross the medium transversely ensures that it is not absorbed significantly during propagation and will be considered constant in the following discussion. Secondly, the phase matching condition is not satisfied for the field generated by four wave mixing in the $2 \vec{k_{\pi}} - \vec{k_{\sigma}}$ direction, that can spoil the transparency. We determine in the following the susceptibility of the atomic system driven by the control field and calculate the group velocity for the probe.

The evolution of the density matrix $\rho$ in the interaction picture leads to the following equations where we have defined the ground and excited populations $n_g= \rho_{aa}+\rho_{bb}, n_e= \rho_{cc}+\rho_{dd}=1-n_g$, the coherences $\rho_{\pi}= \rho_{ca}-\rho_{db}, \rho_{\sigma}= \rho_{da}+\rho_{cb}$ responsible for the $\pi$ and $\sigma$ polarized radiated fields respectively and the imaginary parts of the ground and excited Zeeman coherences $\rho_{zg}=2i \text{Im}\left(\rho_{ab}\right), \rho_{ze}=2i \text{Im}\left(\rho_{cd}\right) $:

\begin{eqnarray}
\label{eq1}
i\dot{n}_g =2i\text{Im} \left(\fpi \rpis +\fsig \phsm \rsigs \right) +i n_e \gmm
\\
\label{eq2}
i \dot{\rho}_{\pi} = \fpi\left(n_e-n_g\right)+\fsig\phsm\left(\rzg+\rze\right)+\delpb^{*}\rpi
\\
\label{eq3}
i\dot{\rho}_{\sigma} = - \fpi\left(\rzg+\rze\right)+\fsig\phsm\left(n_e-n_g\right)+\delpb^{*} \rsig
\\
\label{eq4}
i\dot{\rho}_{zg} = 2\text{Re}\left(\fsig\phsm\rpis-\fpi\rsigs\right)
\\
\label{eq5}
i\dot{\rho}_{ze} = i\dot{\rho}_{zg} -i\gmmz \rze
\end{eqnarray}

Here $\fpi=\frac{d\epsilon_{\pi}}{\hbar},\fsig=-\frac{d\epsilon_{\sigma}}{\hbar}; \left(\left|\fsig\right|<< \left|\fpi\right|\right)$ are the Rabi frequencies for the control and the probe fields  with $d=\left\langle a\right|\vec{D}\cdot\vec{e_{\pi}}\left|c\right\rangle $ ($\vec{D}$ is dipole moment); $\delpb=\delp +i\gmmd$ where $\delp=\omega_0-\omega_{\pi}$ is the detuning of the control field and $\gmmd$ is the rate at which $\sigma$ and $\pi$ coherences relax ; $ \Phi=\Delta t - \left(\vec{k_{\sigma}}-\vec{k_{\pi}}\right)\cdot\vec{r}$ where $\Delta=\left(\omega_{\sigma}-\omega_{\pi}\right)$ is the dephasing between the control and the probe fields and $\gmm,\gmmz$ are the relaxation rates for the populations and the excited zeeman coherence. In the absence of non-radiative homogeneous processes, the rates $\left(\gmmz,\gmmd\right)$ reduce to $\left(\gmm,\gmm/2\right)$. We are interested next in the behaviour of $\rsig$. The Hamiltonian and the total polarization are modulated in space and time which allows us to expand the density matrix in a Floquet development as $\rho=\sum_{n=-\infty}^{\infty} \rho^{\left(n\right)} e^{-in\Phi}$. The coherence that radiates in the direction of the probe field is $\rsigm$ and $\rho_{\sigma}^{\left(-1\right)}$ doesn't radiate because of the phase matching considerations as already discussed. For the weak probe field $\left(\vert\fsig\vert << \sqrt{\gmm\gmmd}\right)$ and at first order with respect to the amplitude of the probe, the stationary solution for $\rsigm$ can be derived from Eq.~(\ref{eq3}) as:
\begin{equation}
	\label{eq6}
	\rsigm = \left(\delpb^{*}-\Delta\right)^{-1} \left[\fsig \left(\nge \right) +\fpi\left(\rzo\right) \right]
\end{equation}
Here, $\nge=\frac{\left|\delpb \right|^2}{4\fpi^2\gmmd\gmm^{-1}+\left|\delpb \right|^2}$ is the static part of the population difference and the first term in Eq.~(\ref{eq6})shows the absorption of the probe by this population. This absorption is compensated by the second term which represents the scattering of the control field off the Zeeman spatial-temporal grating. This second term also represents the cross-Kerr effect \cite{ref14}. In the stationary regime we have:
\begin{equation}
\label{eq7}
	\rzo=-\frac{\fsig}{\fpi}\left(\nge\right)
		\left[
					1-\Delta \left(\delpb^{*}-\Delta\right) W\left(\Delta,\delpb\right)
		\right]
\end{equation}
with
\begin{equation}
\label{eqW}
W= 
				\frac	{\left[\fpi^2 M + \delpb \left(\Delta +\delpb\right)\right] \delpb^{-1}}
							{2\fpi^2 M \left(\Delta+i\gmmd\right)+\Delta\left(\delpb^{*}-\Delta\right)\left(\Delta+\delpb\right)}
\end{equation}
and $M= \frac{2\Delta+i\gmmz}{\Delta+i\gmmz}$. We see from Eq.~(\ref{eq6},\ref{eq7}) that when the two fields have the same frequency $\left(\Delta=0\right)$, the compensation of the absorption of the probe is perfect. This creates a transparency window in the spectral absorption profile of the probe field. This method of producing transparency is in contrast with EIT in a degenerated lambda system for which atoms are pumped into a dark state. Indeed, we can identify two $\Lambda$ systems $\left(\left|a\right\rangle,\left|b\right\rangle,\left|c\right\rangle\right)$ and $\left(\left|a\right\rangle,\left|b\right\rangle,\left|d\right\rangle\right)$ in our duplicated two level system and define dark sates for each as $\left|D\right\rangle=-\fsig e^{-i\Phi} \left| a \right\rangle +\fpi \left| b\right\rangle $ and $\left|\acute{D}\right\rangle=\fpi \left| a \right\rangle +\fsig e^{-i\Phi} \left| b\right\rangle $ respectively. It can be seen that when the matching condition $\left(\fsig e^{-i\Phi}\right)^2=-\fpi^2$ is satisfied, the two $\Lambda$   systems share a common dark state making CPT possible. This situation has been identified and studied in four level systems by many groups \cite{ref7,ref8,ref9,ref10,ref11}. In our situation, the control field is much stronger than the probe $\left(\left|\fpi\right|>> \left|\fsig\right|\right)$ and $\Phi$ is spatially varying, thus ruling out the possibility of CPT. 

\begin{figure}
\includegraphics[width=3in]{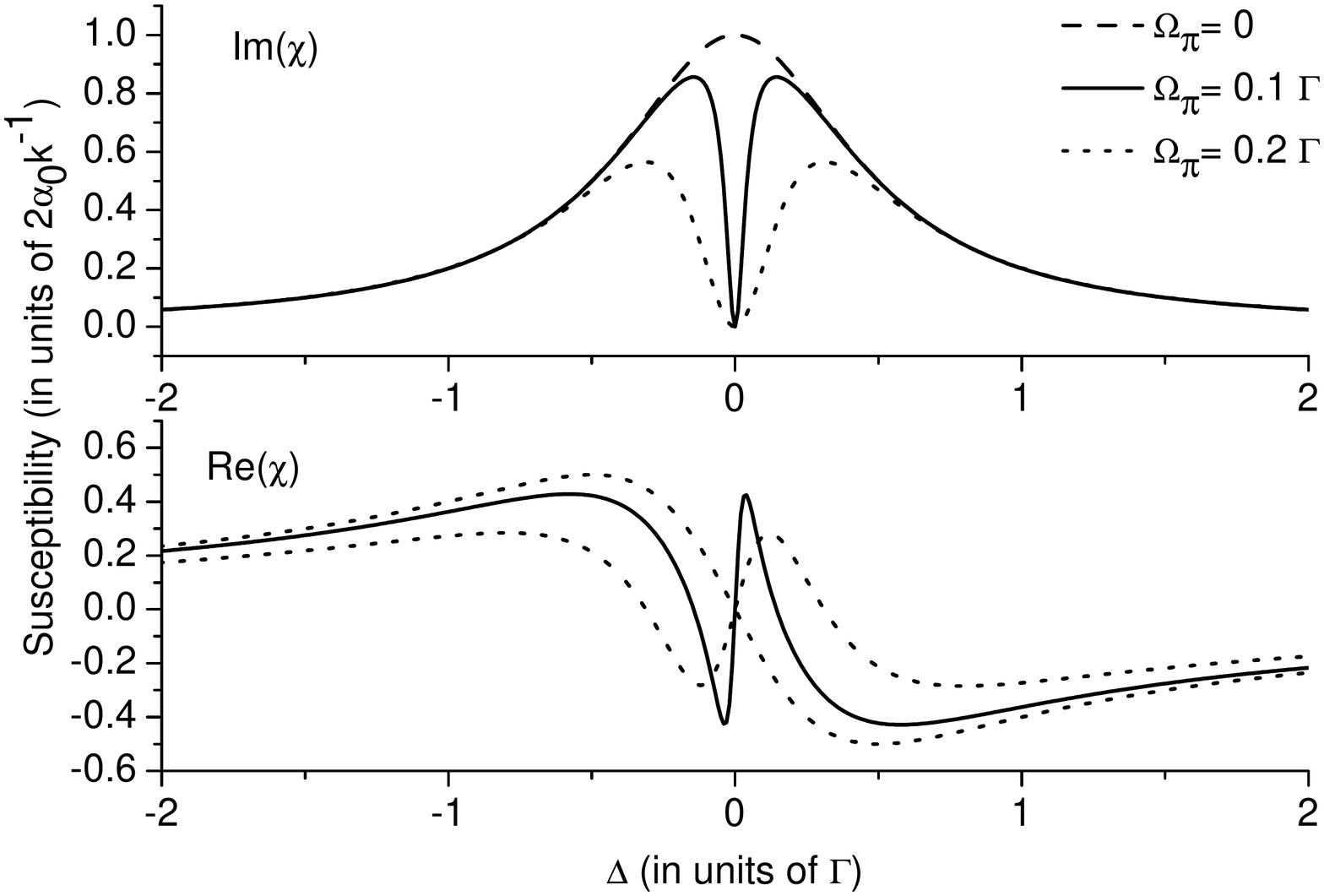}
\caption{\label{fig2} : Profiles of the Real and Imaginary parts of the susceptibility for the probe beam for different strengths of the control beam. The others parameters are $\delp=0,\gmmd=\gmm/2,\gmmz=\gmm$.}
\end{figure}

The effective susceptibility of the system for the probe field is $\chi=\frac{2\alpha_0 \gmmd}{k}\frac{\rsigm}{\fsig}$. Using Eq.~(\ref{eq6},\ref{eq7}) we get:
\begin{equation}
	\label{eq8}
	\chi\left(\Delta,\delp\right) =\Delta
	\frac{2\alpha_0\gmmd}{k}\left(\nge\right)W\left(\Delta,\delpb\right)
\end{equation}
Here $k=\omega_0/c$ and $\alpha_0$ is the field absorption coefficient at resonance given by the relation $\alpha_0=\frac{Nd^2\omega_0}{2c\hbar\epsilon_0\gmmd}$ where $N$ is atomic density. Fig.\ \ref{fig2} gives gives the behaviour of the real and imaginary part of the susceptibility as the function of the detuning $\Delta$ for different strengths of the control field. Two important features can be noted. First, the absorption profile $\text{Im}\left(\chi\right)$ exhibits a dip with a minimum $0$ for $\Delta=0$, as has been discussed before. Secondly, the width of the dip decreases as the control intensity is decreased. Correspondingly the dispersion profile $\text{Re}\left(\chi\right)$ exhibits a very abrupt variation around $\Delta=0$ which is responsible for the slow light. A simplified expression for the absorption profile can be worked out in the limit when the detunings and the control field strength are small as compared to relaxation rates. For $\fpi,\delp,\Delta << \gmmd,\gmmz$, we have $\text{Im}\left(\chi\right)\approx\frac{\alpha_0}{k} \frac{\gmmd^2\Delta^2/2\fpi^4} {1+\left(\gmmd^2\Delta^2/4\fpi^4\right)}$. This is an inverted Lorentzian with the width $~4\fpi^2\gmmd^{-1}$. The relative width of the transparency window compared with the absorption profile (linewidth $2\gmmd$)  is thus $2\left(\fpi/\gmmd\right)^2$  and can be reduced significantly by decreasing the control field intensity. This intensity dependence of the transparency window is similar to that in EIT method \cite{ref1} . In contrast, in CPO technique, the width of the spectral hole is at minimum given by the population relaxation rate and the dip can not be arbitrary reduced \cite{ref4,ref12}. For a probe with a small frequency spread across $\Delta$ , the group velocity is $v_g=c\left( 1+ \frac{\text{Re}\left(\chi\right)}{2}+\frac{\omega_0}{2}\frac{\partial\text{Re}\left(\chi\right)}{\partial\Delta}
\right)_{\Delta=0}^{-1}$ Using Eq.~(\ref{eq8}), we find that
\begin{eqnarray}
	\label{eq9}
	v_g = c \left(
									1+ \frac{c\alpha_0\gmmd}{2\fpi^2}g\left(\fpi,\delpb\right)	
								\right)^{-1} 	
\end{eqnarray} 
with $g\left(\fpi,\delpb\right)= \frac{\left|\delpb\right|^2-\fpi^2}{4\fpi^2\gmmd\gmm^{-1}+\left|\delpb\right|^2}$. For weak intensities of the control field such as $\left|\fpi\right| << \gmmd$, we have $g \approx 1$ and Eq.~(\ref{eq9}) simplifies to the expression that is obtained in EIT methods \cite{ref1}. Very small group velocities can be reached by decreasing the control field intensity and/or increasing the atomic density of the medium. Using $\vert\fpi\vert=\left|d/\hbar\right|\sqrt{I_{\pi}/2c\epsilon_0}$, where $I_{\pi}$ is the control field intensity, we obtain $v_g \approx 2I_{\pi}/N\hbar\omega_0$. For $I_{\pi}=1 mW/cm^3$ and $N=10^{12} at/cm^2$, we get $v_g \approx 60 m/s$. The group delay $\tau=L\left(v_g^{-1}-c^{-1}\right)$ with $L$ the length of the medium, can be approximated as $\frac{\alpha_0 L}{2\fpi^2}\gmmd$. The figure of merit is $\tau/T$ with $T$ the pulse duration. Since $\fpi^2\gmmd^{-1}/\sqrt{\alpha_0 L}$ represents the width of the transparency window for the \textit{probe spectrum}, we need to have $T>> \sqrt{\alpha_0 L}\fpi^{-2}\gmmd$ to ensure that the entire probe spectrum is located within this window. The figure of merit is important only if the optical depth $\alpha_0 L$ is large enough to compensate for this effect. 

The main limitation of our method originates from higher order non-linear effects we have neglected so far. The transparency is related to the vanishing absorption around $\Delta=0$. However, this is true only at the first order with respect to the probe amplitude. Higher order terms may spoil this transparency. The coherence $\rsigm$ can be evaluated at the next order of probe field amplitude (at $\Delta=0$ ) as:
\begin{eqnarray}
	\label{eq10}
	\rsigm = \frac{\fsig \left(\delpb^{*}\right)^{-1} \left(\fsig/\fpi\right)^2}
								{\left(1+ 4 \fpi^2\gmmd\gmm^{-1}\left|\delpb\right|^{-2}\right)^2}
\end{eqnarray}
This contribution is very small for $\vert\fsig\vert<<\vert\fpi\vert$, however, as the probe field obeys the propagation equation \cite{ref3,ref4,ref5,ref6,ref7,ref8,ref9,ref10,ref11,ref12} $\frac{\partial \fsig}{\partial y}+\frac{1}{c}\frac{\partial \fsig}{\partial t} = i \alpha_0 \gmmd \rsigm $, this small contribution can be amplified during propagation and can spoil transparency.  This requires that an additional condition $X=\alpha_0 L \left|\fsig/\fpi\right|^2<<1$  be fulfilled in order to obtain a deep transparency window.  

\begin{figure}
\includegraphics[width=3in]{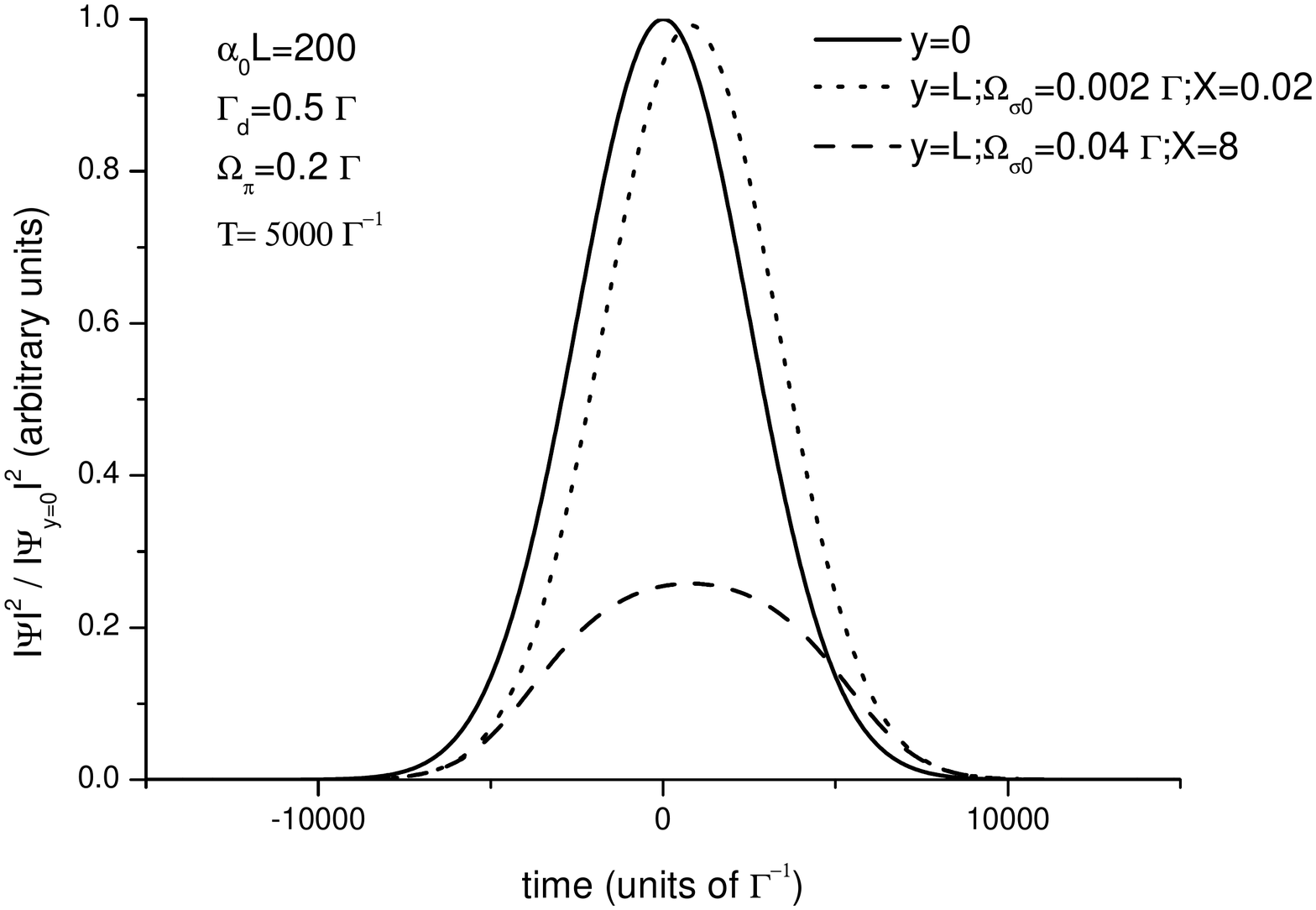}
\caption{\label{fig3} Influence of non-linear effects. Temporal profile of the dark polariton intensity for different values of $X$. We have $\fsig\left(t,y=0\right)=\Omega_{\sigma 0} e^{-\left(t/T\right)^2}/\sqrt{\pi}$ and $\fpi=\text{cte}$. Dotted line: transparency is ensured and the polariton is delayed in time. Dashed line: the polariton is damped and distorted.}
\end{figure}

An important feature of slow light by EIT is the presence of dark polariton that couples the propagating beam and the atomic coherences \cite{ref5}. We can define a dark polariton as $\psi\left(y,t\right)=\cos\left(\theta\right) \fsig \left(y,t\right)- \sin\left(\theta\right)\sqrt{\kappa}\left(\rzo\right)$  with $\kappa=\frac{c\alpha_0\gmmd}{2}g\left(\fpi,\delpb\right)$  and $\tan\left(\theta\right)=\sqrt{\kappa}/\fpi$  . In the situation where transparency for the entire probe spectrum is ensured, Eq.~(\ref{eq7}) still holds and the instantaneous Zeeman coherence can be approximated as $\rzo \approx -\fsig/\fpi$. It propagates within the material along with the probe. The dark polariton then evolves according to the (shape preserving) equation $\partial\psi/\partial t + c \cos^2\left(\theta\right) \partial \psi/\partial y =0$. The important difference with dark polaritons in $\Lambda$ system \cite{ref5} is that here only a part of the Zeeman coherence constitutes the bound part of the polariton. When higher order terms given by Eq.~(\ref{eq10}) become important, the polariton is modified during propagation, experiencing both damping and shape distortion. This may happens if we decrease the intensity of the control field or if the probe pulse is not enough weak. Fig.\ \ref{fig3} shows the damping of the dark polariton in a situation where the perfect transparency is not realized in contrast to the temporally delayed polariton obtained for perfect transparency (signifying slow light). For the chosen parameters, the dominant contribution for the polariton is the bound material part and is damped when transparency is not ensured during propagation. This effect represents an additional limitation for storing light using this method. 

In conclusion, we have introduced a new method based on Zeeman Coherence Oscillation (CZO) for slowing light in a duplicated two-level system. The transparency for the probe field is achieved by a balance between the medium absorption of the probe and the control field diffraction by the Zeeman coherence grating. Like EIT, the width of the transparency window is determined by control field intensity. The transparency is ensured for $\alpha_0 L \left|\fsig/\fpi\right|^2 << 1$ whereas efficient reduction of light velocity requires $\fpi^2 << c \alpha_0 \gmmd$. This indicates that this method can work only for small flux of the probe pulse. The results presented here lead us also to conclude that EIT, CPO and CZO can be described as different manifestations of the wave mixing induced by the two exciting field. The transparency results from a balance between the absorption of the probe and diffraction of the control field off some grating. Indeed, even EIT method can be described in this way for a degenerate $\Lambda$ system and results from a scattering process in which the control beam is diffracted by the ground level Zeeman coherences \cite{ref3,ref10}. This feature is generally overlooked in the literature because of the more powerful description based on the trapping state, but it can be the key idea for implementation of slow light in more complex systems where no dark state exists.
\begin{acknowledgments}
We gratefully acknowledge J-L Le Gouët and B. Macke for truly enlightening discussions and useful criticism of this work.
\end{acknowledgments}


\end{document}